# Flying V and Reference Aircraft Evacuation Simulation and Comparison

J. Gebauer[1] and J. Benad[2]

*Technische Universität Berlin, 10623 Berlin, Germany*

**A preliminary comparison of evacuation times of the Flying V and the Airbus A350-900 is presented in this study. A simple simulation tool based on the technique of cellular automata was created to model the evacuation process for different closed door configurations. Certification regulations state that the time to evacuate a civil aircraft in case of an emergency with half of all exit doors closed must be less than 90 seconds. The results of this study indicate that the shorter V shaped cabin has advantages over the longer conventional reference cabin for cases when passengers need to evacuate towards the front or the back of the aircraft. Disadvantages occur when the passengers in the V shaped cabin need to evacuate more towards one side (left or right wing) of the aircraft. A more detailed simulation model to further investigate these cases is currently created by the authors.**

## 1. Introduction

The Flying V is a new configuration for an efficient aircraft. Conceived and studied by Benad from 2013 to 2015 [1-4], the concept is currently further developed and researched in a project led by Vos at Delft University of Technology [5-10] in collaboration with several partners. With passenger compartments arranged in the shape of a V, the cabin geometry of the airplane differs significantly from the cabin geometry of the conventional tube and wing configuration (see Figure 1). Certification regulations state that the evacuation time of a civil passenger aircraft must not exceed 90 seconds when half of all doors are closed (CS-25, [11]). For flying wings in particular, this requirement has always been a topic of some concern [12,13]. There are various techniques to simulate emergency evacuations, among them are cellular automaton models [14], or the continuous social force model [15]. An exemplary evacuation software is airExodus [16].

For a first preliminary analysis of the evacuation of the Flying V, a simulation tool was developed by the authors based on the technique of cellular automata with a floor field model (see [14]). Here, a discrete domain is introduced, where each cell state can be empty ("zero") or occupied ("one"). A passenger then decides where to go by a probability calculated by layering different fields. In the present study, a single parameter $r$ characterizing the level of random motion of the passengers during the evacuation process is introduced and calibrated to match evacuation times of existing airplane configurations. With the calibrated tool, multiple simulations are executed to compare the evacuation times of the Flying V and the Airbus A350-900 reference aircraft for different closed door configurations.

## 2. Method

In order to create a simple simulation tool, cellular automatons are used. A cellular automaton is a discrete model that is able to map complex processes by setting simple rules. Hereby, the observed domain is the passenger cabin that is modeled by a grid consisting of square cells. The dimension for one cell was chosen with 40 cm × 40 cm to model dimensions of seats and aisles as well as the space a pedestrian occupies (see [17,18]). The generated grids are shown in Figure 1.

The walls and seats receive the cell state "one", which is permanent over the course of the simulation. However, the cell state for passengers changes over time. The assumption was made that only the closest adjacent cells have an impact on a passenger, so that a Moore neighborhood with a radius of one was chosen. In addition to the neighborhood, the collision of passengers needs to be specified more precisely. A single cell can only be occupied by one passenger in one time step. When multiple passengers have the same target cell, one passenger is chosen randomly. This passenger is allowed to move to this target cell while the other passengers are prohibited from moving at all. In order to define when a passenger can be seen as evacuated, boundary conditions need to be set. When a passenger enters an exit door, this passenger is considered evacuated and is ignored in the next time step. In addition, the transition between the legs of the Flying V is crucial. In the present preliminary model, two separate latices are aligned with each leg of the V. Where both legs meet, transition conditions are applied. A more refined model where the entire Flying V geometry is modelled with a single lattice with a higher resolution is currently under development by the authors.

---

[1] MSc. student, Institute of Mechanics
[2] Research assistant, Institute of Mechanics, j.benad@tu-berlin.de

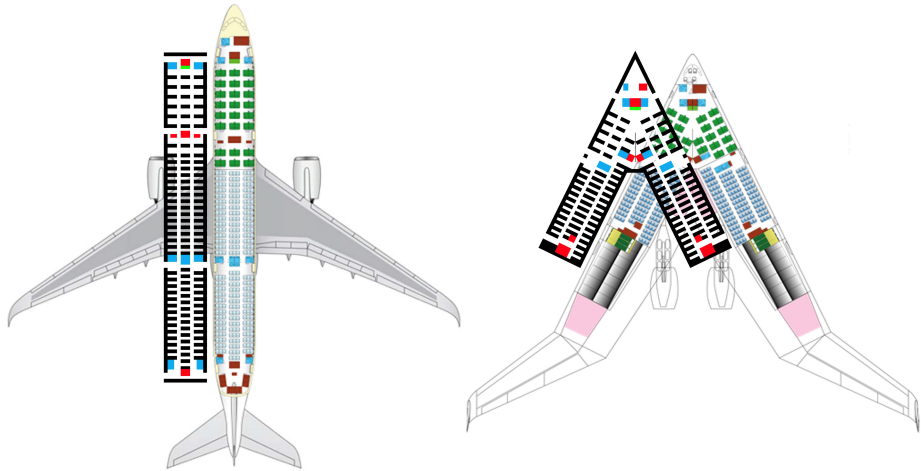

Figure 1: The cabin geometry of the Airbus A350-900 (left) and Flying V (right) modeled with a grid based on square cells.

The movement of a passenger depends on a transition probability *p*. In accordance with the floor field model [14], this probability is calculated by layering different fields. Three different fields are taken into account: a gradient, distance and direction field. The gradient field presents the urge of each passenger to reach the exit doors with the shortest way possible. The distance field adds an entirely random motion to the passengers. In the present model, its influence decreases linearly with the distance to each exit. The overall influence of the distance field can be adjusted by the single parameter *r*. Therefore, in the present study, this single parameter *r* is used to characterize the level of random motion of the passengers during the evacuation process. This parameter can be calibrated to match evacuation times of existing airplane configurations. Additionally, a correction field is applied to specific small areas with influence parameter *q* to guarantee that passengers do not get stuck in a dead end. The transition probability used in this simulation is

$$p_{ij} = \left[\left(r\, p_{d_{ij}} + (1-r)p_{g_{ij}}\right)q + (1-q)p_{c_{ij}}\right](1 - w_{ij}) \qquad (1)$$

with $i, j \in \{1, 2, 3\}$, where $p_{d_{ij}}$ represents the distance field, $p_{g_{ij}}$ the gradient field, $p_{c_{ij}}$ the correction field, and $w_{ij}$ the wall grid, where a movement is prohibited. The variables *i* and *j* represent the adjacent cells that are considered for the calculation of the probability.

In the preliminary model, all passengers will be moving with the same velocity of $v \approx 1.3$ m/s, which is the average velocity for a pedestrian [18]. In the present simulation, where a passenger walks with one cell per time step, this translates to a time step of approximately 0.3 s.

In order to calibrate the simulation tool for the Airbus A350-900, data from trials or other evacuation models was researched. No values were found for the reference aircraft, but due to similarity in exit door arrangements and seat capacity, values presented in [19] obtained from a simulation with airExodus applied to the Boeing 767 were used to calibrate the present preliminary model. From this study, a target evacuation time of 60 s could be derived for a case where all doors on the right side of the aircraft are closed. This time excludes the response time of crew members. Multiple simulations were run for the calibration. The results of these simulations are shown in Figure 2. Based on the outcome of these simulations, the parameter was set to $r = 10^{-2}$.

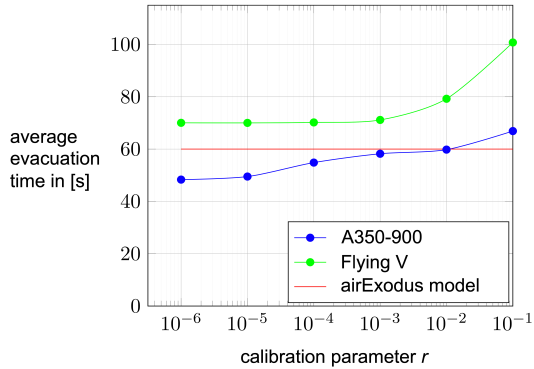

Figure 2: Multiple simulation were executed with parameter *r*, characterizing the level of random motion of the passengers, ranging from $10^{-6}$ (almost no random motion) to $10^{-1}$ (more random motion). The value was calibrated for the A350-900 to match the Air Exodus model (red).

# 3. Results

An exemplary simulation with closed doors on the right side of the Flying V and the reference is shown in Figure 3. During the first couple of time steps, the passengers leave both airplanes efficiently. Subsequently, passengers in the business class exit quickly while jam formations occur in the economy class areas of both airplanes.

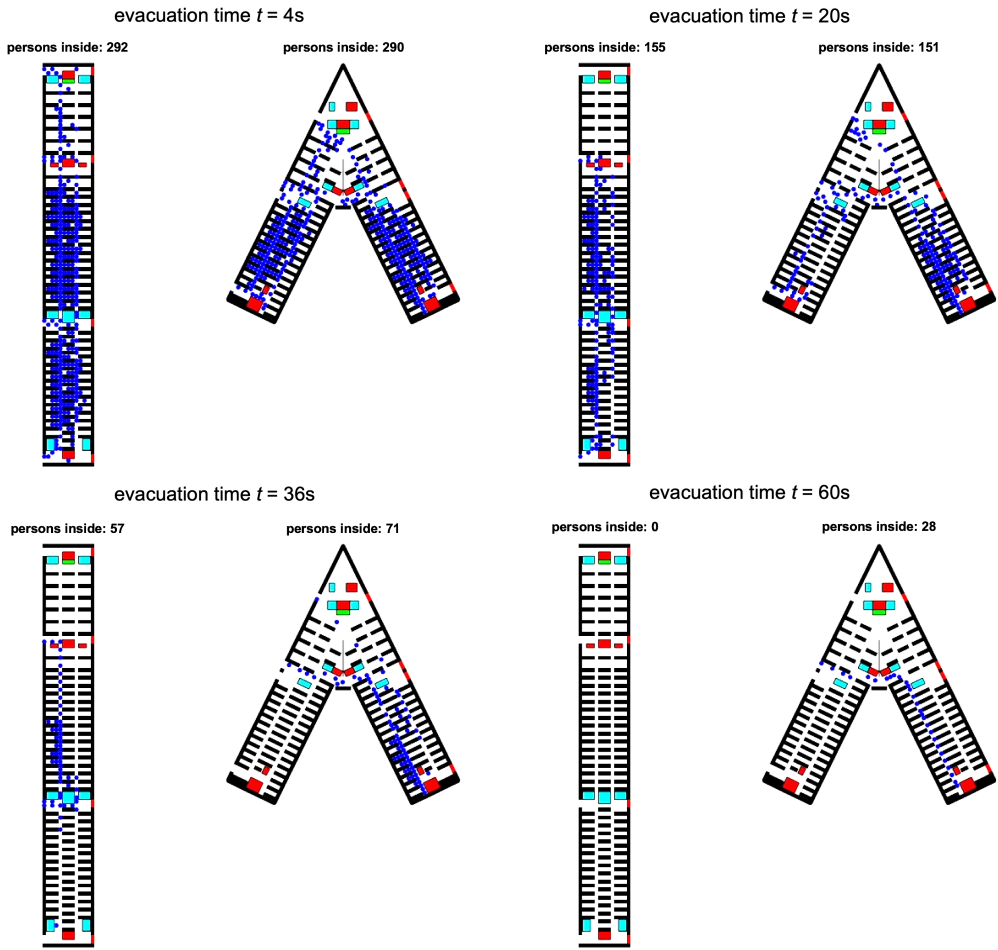

**Figure 3: An exemplary evacuation simulation with closed doors on the right side of the Flying V the reference. The blue dots represent the passengers. Results are displayed for a parameter $r = 10^{-2}$.**

Figure 4 displays the number of passengers inside the Flying V and the reference over the evacuation time during one exemplary simulation process where the doors on the right side of both planes are closed. The different colors of the curve sets in each diagram correspond to the parameter $r$. At the beginning of the simulation an almost linear progression occurs which is similar for both aircrafts. A significant difference occurs when approximately 80 passengers are left in the cabin. From that moment on, the evacuation of the Flying V proceeds slower.

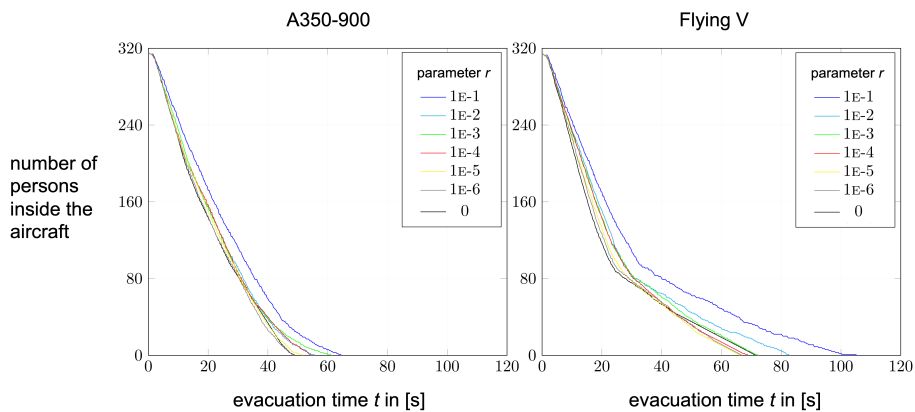

**Figure 4: Number of passengers inside Flying V and the reference over the evacuation time during one exemplary simulation process where the doors on the right side of the planes are closed. Colored curve sets are displayed for different parameter $r$ ranging from $r = 0$ to $r = 10^{-1}$.**

Various closed door configurations were examined for the Flying V and the reference aircraft. They are displayed in Figure 5. It shall be emphasized, that these are preliminary results obtained with an extremely simple tool. One should exert great caution with these results, especially with the quantitative values. Nevertheless, the results do indicate that the shorter V shaped cabin has some advantages over the longer tube cabin if the evacuation takes place only towards the front or only towards the rear of the aircraft (cases 5 and 6). For example, the preliminary tool showed a reduction in evacuation time of 62% for the Flying V when compared to the reference when half of all doors in the front of the aircraft are closed (case 5). When half of all doors in the back of the aircraft are closed (case 6), a reduction of 34% in evacuation time was obtained. This seems to indicate a similar trend as was obtained in a recent study [20] where boarding times of the Flying V and the A350-900 were simulated using agent based modelling. In this study, a reduction of 30% in boarding time was obtained for the Flying V where passengers can proceed from the front to the back of the aircraft using four available aisles as opposed to two aisles in the reference aircraft. Results of the present study obtained for case 2 displayed in Figure 5 indicate that disadvantages in the evacuation process may occur when the passengers in the V shaped cabin need to evacuate solely towards one side of the aircraft. In this case, an increase of 37% in evacuation time for the Flying V was obtained when compared to the reference with the preliminary tool.

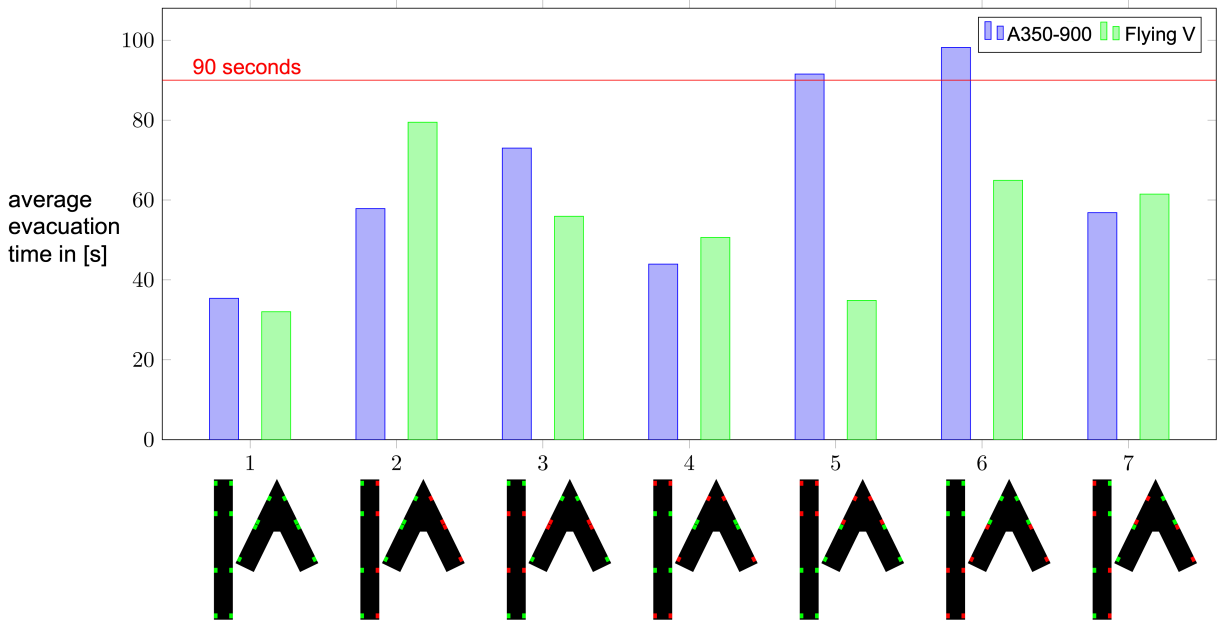

**Figure 5: Average evacuation times for the Flying V and the reference aircraft displayed for various closed door configurations and a parameter $r = 10^{-2}$. Note that the displayed times are only of the evacuation process and exclude crew reaction times at the beginning of the evacuation. Note further, that these are preliminary results obtained with an extremely simple tool. One should exert great caution with these results, especially with the quantitative values.**

## 4. Conclusion

A preliminary analysis of the evacuation of the Flying V and the A350 reference aircraft was presented in this study. A simple simulation tool based on the technique of cellular automata was created to model the evacuation process. Therein, a passenger decides where to go by a probability calculated by layering different fields. A single parameter $r$ characterizing the level of random motion of the passengers during the evacuation process was introduced and calibrated to match evacuation times of existing airplane configurations. With the calibrated tool, multiple simulations were executed to compare the evacuation times of the Flying V and the Airbus A350-900 reference aircraft for different closed door configurations. This comparison is displayed in Figure 5. The results indicate that the V shaped cabin has some advantages over the conventional cabin when passengers are evacuated solely towards the front, or solely towards the rear of the aircraft. Disadvantages occur when passengers are evacuated towards one side of the aircraft.

It shall be emphasized, that these are preliminary results obtained with an extremely simple tool. One should exert great caution with these results, especially with the quantitative values. A more detailed simulation model to further investigate the topic and substantiate the results given in this preliminary study is currently under development by the authors. For instance, in the present study two separate very rough lattices are aligned with each leg of the V. Where both legs meet, transition conditions are applied. In future studies by the authors, a more refined model where the entire Flying V geometry is modelled with a single lattice with a higher resolution will be used. This will lead to a more natural motion of the passengers and detailed features of the cabin geometry can then be taken into account. Furthermore, in the present model, a passenger may stand in line with many others and wait to get out of the closest door rather than to look for another entirely free door which is only a short distance further away. Such details will also be addressed in future simulation tools.